# In Pursuit of Spreadsheet Excellence

Grenville J. Croll
Chair
European Spreadsheet Risks Interest Group
www.eusprig.org
Grenville@spreadsheetrisks.com

**Abstract**

**The first fully-documented study into the quantitative impact of errors in operational spreadsheets identified an interesting anomaly. One of the five participating organisations involved in the study contributed a set of five spreadsheets of such quality that they set the organisation apart in a statistical sense. This virtuoso performance gave rise to a simple sampling test – The Clean Sheet Test - which can be used to objectively evaluate if an organisation is in control of the spreadsheets it is using in important processes such as financial reporting**.

## 1. Introduction

Following a protracted search, Powell, Lawson and Baker [Powell et al, 2007] managed to find five organisations which were willing to contribute operational spreadsheets for detailed examination. The objective of their study was to determine the financial impact of spreadsheet errors in organisations.

The five organisations comprised two consulting companies, a very large financial services firm, a manufacturing company and a college. Each organisation provided five differing spreadsheets for examination by the researchers. Amongst other pre-specified criteria, the volunteered spreadsheets were quite large, well understood by the volunteers and developed by the volunteers within the last twelve months. Each spreadsheet was then independently examined by two researchers and the issues determined by the researchers' examination were pooled. The agreed issues observed jointly by the researchers were fed back to the volunteer developers and categorised as an error, poor practice or not an error. Actual errors were then corrected and the change in the relevant output cell was recorded as the quantitative measure of the impact of the error.

## 2. Original Tabulated Results

We reproduce, in a slightly condensed form, Powell et al's Table 1 which summarises their results. As is evident, following the detailed examination of 25 spreadsheets contributed by five organisations, they found 381 Issues which translated into 177 agreed errors. 79 of the errors had a non zero impact. The absolute impact of the maximum error found in each workbook varied from $0.22m to $110m dollars. The total of the maximum absolute error impacts across the original 25 worksheets is in excess of $259m.

**Table 1 (of Powell et al 2007, condensed)**

| Organisation/ Workbook | #Issues | #Errors | Errors with Non-zero Impact | Max % Impact | Max Abs Impact |
|---|---|---|---|---|---|
| 1.1 | 7 | 3 | 0 | | |
| 1.2 | 50 | 6 | 5 | 28% | $32m |
| 1.3 | 18 | 7 | 3 | 137% | $110m |
| 1.4 | 4 | 1 | 0 | | |
| 1.5 | 0 | 0 | 0 | | |
| 2.1 | 19 | 6 | 5 | 3.6% | $14m |
| 2.2 | 27 | 11 | 7 | 16% | $74m |
| 2.3 | 6 | 0 | 0 | | |
| 2.4 | 30 | 4 | 3 | 416% | $11m |
| 2.5 | 40 | 2 | 2 | | |
| 3.1 | 19 | 2 | 2 | 5.3% | $0.24m |
| 3.2 | 1 | 1 | 0 | | |
| 3.3 | 11 | 2 | 2 | 16% | $4.9m |
| 3.4 | 6 | 1 | 0 | | |
| 3.5 | 23 | 1 | 0 | | |
| 4.1 | 27 | 22 | 12 | 117% | $13m |
| 4.2 | 8 | 4 | 2 | 142% | $0.27m |
| 4.3 | 0 | 0 | 0 | | |
| 4.4 | 1 | 0 | 0 | | |
| 4.5 | 79 | 44 | 27 | 39% | $0.22m |
| 5.1 | 2 | 0 | 0 | | |
| 5.2 | 2 | 0 | 0 | | |
| 5.3 | 0 | 0 | 0 | | |
| 5.4 | 0 | 0 | 0 | | |
| 5.5 | 1 | 0 | 0 | | |
| Total | 381 | 117 | 70 | | |

Note that in organisation five, the number of issues found was very low compared to the other four organisations. Note also that the number of resulting errors in each of their five workbooks was nil. Given the ubiquity of spreadsheet error as reported in every other study [Panko, 2000, 2007], the existence of one organisation producing five error free workbooks out of five volunteered for examination appears at the outset to be anomalous.

Powell et al recognise this and state that:

> *"Organisation 5 is a small consulting company with highly educated employees and a culture that demands excellence. The spreadsheets we audited from this firm were works of art: thoughtfully designed, well documented, and error free"*

Our submission is that the results from this firm are anomalous in that such excellence is presently not the norm. We hypothesise that the spreadsheets produced by this organisation for this study are the result of a significant software engineering process.

The development of spreadsheets is an error prone process [Panko, 2000] [Panko & Ordway, 2005]. The only known method of systematically detecting and correcting spreadsheet errors is through repeated cell by cell inspection of every cell in a spreadsheet by multiple testers [Panko, 2006]. It is not possible to produce by chance a large error free spreadsheet. Field and laboratory testing over the last twenty years reflects this. The production of error free spreadsheets implies the use of repeated cell by cell testing by multiple testers or some other noteworthy but hitherto unknown software engineering or other process [Panko, 2007].

**3. Anomaly Detection**

We can test statistically the likelihood that an organisation might produce five error or defect free spreadsheets by chance alone.

If we look at column 3 of Table 1, we see a column entitled "Errors". From this column we can create a further column entitled "Errors?". We can represent an error free spreadsheet with a 0, and a spreadsheet with at least one error with a 1.

If we look at column 4 of Table 1, we see a column entitled "Errors with Non Zero Impact". From this column we can create a further column entitled "Defects?". We can represent a defect free spreadsheet with a 0, and a spreadsheet with at least one defect with a 1.

We can also create a final column which represents our hypothesis. The final column is 0 for each of the spreadsheets from organisations 1 to 4 representing our hypothesis that we do not believe them to be anomalous. These organisations exhibit the characteristics found in all known previous field testing. Each of the 5 spreadsheets from organisation 5 is represented by a 1 in the final column representing our hypothesis that they are all anomalous in a singular way.

We reproduce this additional information in Table 2 below.

We can now use the simple statistical technique of Linear Regression to determine whether there is any statistical correlation between the hypothesised anomalous status of five spreadsheets and their Error and Defect status. We later confirm this result using Logistic Regression.

**Table 2 – Errors & Defects**

| Organisation/ Workbook | #Errors | Errors with Non-zero Impact | Errors? | Defects? | Anomalous? |
|---|---|---|---|---|---|
| 1.1 | 3 | 0 | 1 | 0 | 0 |
| 1.2 | 6 | 5 | 1 | 1 | 0 |
| 1.3 | 7 | 3 | 1 | 1 | 0 |
| 1.4 | 1 | 0 | 1 | 0 | 0 |
| 1.5 | 0 | 0 | 0 | 0 | 0 |
| 2.1 | 6 | 5 | 1 | 1 | 0 |
| 2.2 | 11 | 7 | 1 | 1 | 0 |
| 2.3 | 0 | 0 | 0 | 0 | 0 |
| 2.4 | 4 | 3 | 1 | 1 | 0 |
| 2.5 | 2 | 2 | 1 | 1 | 0 |
| 3.1 | 2 | 2 | 1 | 1 | 0 |
| 3.2 | 1 | 0 | 1 | 0 | 0 |
| 3.3 | 2 | 2 | 1 | 1 | 0 |
| 3.4 | 1 | 0 | 1 | 0 | 0 |
| 3.5 | 1 | 0 | 1 | 0 | 0 |
| 4.1 | 22 | 12 | 1 | 1 | 0 |
| 4.2 | 4 | 2 | 1 | 1 | 0 |
| 4.3 | 0 | 0 | 0 | 0 | 0 |
| 4.4 | 0 | 0 | 0 | 0 | 0 |
| 4.5 | 44 | 27 | 1 | 1 | 0 |
| 5.1 | 0 | 0 | 0 | 0 | 1 |
| 5.2 | 0 | 0 | 0 | 0 | 1 |
| 5.3 | 0 | 0 | 0 | 0 | 1 |
| 5.4 | 0 | 0 | 0 | 0 | 1 |
| 5.5 | 0 | 0 | 0 | 0 | 1 |
| Totals | 117 | 70 | 16 | 11 | 5 |

## 4. Regression Results

A simple linear regression with the Errors? variable as the dependent Y-variable and the Anomalous status as the independent X-Variable shows a reasonable statistical correlation (Adj. $R^2$ = 0.44). The significance statistics for the Anomalous variable show beyond reasonable doubt (t = -4.28, p = 0.00027) that the Errors? status of the five spreadsheets volunteered by organisation five are statistically very different from the other twenty supplied by the other four organisations. The chance that organisation five might have produced such Error Free spreadsheets by chance alone is of the order of 0.03%. It is almost certain therefore that organisation five generated their five spreadsheets by a process that was designed to eliminate error. Note that results from a logistic regression of this binary data are similar [Pezullo, 2008].

The linear regression was repeated using the Defects? column as the dependent Y-variable. The results show a weaker correlation (Adj. $R^2$ = 0.20), however the Anomalous variable is still significant (t = -2.4, p = 0.027). The chance that organisation five should produce five Defect Free spreadsheets by chance is approximately 2.7%.

If organisation five had produced one or two defect free spreadsheets, their performance would have been indistinguishable from the other organisations.

Note that in the 20 spreadsheets that were not anomalous, 16 (75%) of them had errors. This is lower than the usual error rate detected in other studies, where a rate in excess of 90% is normally observed. This may be because error rates are now improving. Or, more likely, due to the use of a repeatable error discovery protocol which may have been absent across other studies, a lack of domain knowledge by the researchers or due to the spreadsheet selection criteria used. Note that in the 20 spreadsheets that were not anomalous 9 (45%) of them were defect free.

**5. The Clean Sheet Test**

Using the data in Table 2 as a guide, if we assume that 25% of all corporate spreadsheets are presently error free, the probability that an organisation might by chance have five error free spreadsheets in a random sample of five is $0.25^5$ or 0.097%. Likewise, the probability that an organisation might by chance have five defect free spreadsheets in a sample of five is $0.45^5$ or 1.8%. Note that approximately 50% of errors are not defects.

These probabilities, together with the size of the defects identified by Powell et al highlight the risks to financial reporting integrity where spreadsheets are used in the financial reporting process.

In addition, the structure of Powell et al's spreadsheet experiment gives us a method for determining if an organisation is in control of its Business Critical spreadsheets.

Firstly, we must randomly select five spreadsheets from an organisation. For example we could select five spreadsheets from the many dozens or hundreds used in the financial reporting process. We should then examine them using Powell et al's process [Powell et al 2006] or another process such as that described in an earlier paper [Croll, 2003].

Secondly, if we find five defect free spreadsheets, then based upon on the error and defect rates detected in this study, the organisation would appear to have a process in place to control errors and defects.

If an organisation is unable to pass the Clean Sheet Test we should be wary about placing any reliance upon any information based upon those spreadsheets or the population from which they have been randomly selected, as Powell et al and other reports [EuSpRIG, 2008] suggest that the Impact of such defects is material, if not critical [Croll, 2005].

**6. The Benefits of Clean Spreadsheets**

There are obvious benefits in ensuring financial integrity through the use of well designed, tested and documented (i.e. Clean or Defect Free) spreadsheets in Financial Reporting and other important corporate processes. The Clean Sheet Test provides a simple method for determining if this is the case for any particular organisation.

Organisations which pass the Clean Sheet Test could be said to be exhibiting Spreadsheet Excellence.

Finally, there is a possibility that a link can be established between Spreadsheet Excellence as determined above and publicly disclosed measures of corporate performance. MacMillan [2000] established a positive statistical link between the use of quantitative decision making methods such as DCF, Monte Carlo Simulation and Real Options and enhanced financial performance within approximately 30 participants in the UK Offshore Oil and Gas Industry. Given that there is a strong relationship between the use of these techniques and the use of spreadsheets, the establishment of a similar link between Spreadsheet Excellence and improved financial performance may not be too difficult to prove and would be a useful and important discovery.

## 7. Conclusion

A consulting organisation delivered a set of five spreadsheets for testing which upon examination turned out to have zero defects. We show that this unusual outcome was statistically measurable and set the organisation apart from the others in the study. We used this observation to create a relatively simple test – The Clean Sheet Test - which can be used to identify Spreadsheet Excellence in organisations. We state the obvious benefits of Spreadsheet Excellence in financial reporting and propose an experiment to determine if there is a statistical link between Spreadsheet Excellence and improved financial performance.

**References**

Croll, G.J. (2003) "A Typical Model Audit Approach", IFIP, Integrity and Internal Control in Information Systems, Vol 124, pp. 213-219, http://arxiv.org/abs/0712.2591

Croll, G.J. (2005) "The Importance and Criticality of Spreadsheets in the City of London", European Spreadsheet Risks Interest Group, http://arxiv.org/abs/0709.4063

EuSpRIG (2008) "Spreadsheet Mistakes: News Stories", http://www.eusprig.org/stories.htm, Accessed 5$^{th}$ June 2008 14:00

MacMillan, F. (2000) "Risk, Uncertainty and Investment Decision-Making in the Upstream Oil and Gas Industry", Ph.D. Thesis, 2000, University of Aberdeen, Scotland.

Pezullo, J.C., (2008) "Logistic Regression", http://statpages.org/logistic.html Accessed 5 Jun 08 10:50

Panko, R., (2000) "Spreadsheet Errors: What We Know. What We Think We Can Do", European Spreadsheet Risks Interest Group, 1st Annual Symposium, University of Greenwich, pp7-18 http://arxiv.org/abs/0802.3457

Panko, R., Ordway, N., (2005) "Sarbanes-Oxley: What About All the Spreadsheets?", European Spreadsheet Risks Interest Group, pp15-47, http://arxiv.org/abs/0804.0797

Panko, R., (2006) "Recommended Practices for Spreadsheet Testing", Proc. European Spreadsheet Risks Interest Group, pp73-84, http://arxiv.org/abs/0712.0109

Panko, R., (2007) "Thinking is Bad: Implications of Human Error Research for Spreadsheet Research and Practice", Proc. European Spreadsheet Risks Interest Group, pp69-80, http://arxiv.org/abs/0801.3114

Powell, S., Baker, K., and Lawson, B. (2006). "An Auditing Protocol for Spreadsheet Models." Spreadsheet Engineering Research Project working paper. http://mba.tuck.dartmouth.edu/spreadsheet/product_pubs.html Accessed 5th June 2008 11:02.

Powell, S., Lawson, B., Baker, K. (2007) "Impact of Errors in Operational Spreadsheets", European Spreadsheet Risks Interest Group, 8th Annual Conference, University of Greenwich, pp57-68. http://arxiv.org/abs/0801.0715